\documentclass[pdflatex,sn-mathphys]{sn-jnl}

\jyear{2022}%
\usepackage{tablefootnote}
\theoremstyle{thmstyleone}%
\theoremstyle{thmstyletwo}%

\theoremstyle{thmstylethree}%

\begin{document}
\title[TRUSformer]{TRUSformer: Improving Prostate Cancer Detection from Micro-Ultrasound Using Attention and Self-Supervision}


\author*[1]{\fnm{Mahdi} \sur{Gilany}}\email{mahdi.gilany@queensu.ca}
\equalcont{These authors contributed equally to this work.}

\author[1]{\fnm{Paul} \sur{Wilson}}
\equalcont{These authors contributed equally to this work.}

\author[1]{\fnm{Andrea} \sur{Perera-Ortega}}
\author[1]{\fnm{Amoon} \sur{Jamzad}}
\author[2]{\fnm{Minh Nguyen Nhat} \sur{To}}
\author[2]{\fnm{Fahimeh} \sur{Fooladgar}}
\author[3]{\fnm{Brian} \sur{Wodlinger}}
\author[2]{\fnm{Purang} \sur{Abolmaesumi}}
\author[1]{\fnm{Parvin} \sur{Mousavi}}

\affil*[1]{\orgdiv{School of Computing}, \orgname{Queen's University}, \orgaddress{\city{Kingston}, \country{Canada}}}

\affil[2]{\orgdiv{Department of Electrical and Computer Engineering}, \orgname{University of British Columbia}, \orgaddress{\city{Vancouver}, \country{Canada}}}

\affil[3]{\orgname{Exact Imaging}, \orgaddress{\city{Markham}, \country{Canada}}}


\abstract{A large body of previous machine learning methods for ultrasound-based prostate cancer detection classify small regions of interest (ROIs) of ultrasound signals that lie within a larger needle trace corresponding to a prostate tissue biopsy (called biopsy core).



These ROI-scale models suffer from weak labeling as histopathology  results available for biopsy cores only approximate the distribution of cancer in the ROIs. ROI-scale models do not take advantage of contextual information that are normally considered by pathologists, i.e. they do not consider information about surrounding tissue and larger-scale trends when identifying cancer. We aim to improve cancer detection by taking a multi-scale, i.e. ROI-scale \emph{and} biopsy core-scale, approach.

\emph{Methods:} Our multi-scale approach combines (i) an ``ROI-scale" model trained using self-supervised learning to extract features from small ROIs and (ii) a ``core-scale" transformer model that processes a collection of extracted features from multiple ROIs in the needle trace region to predict the tissue type of the corresponding core. Attention maps, as a byproduct, allow us to localize cancer at the ROI scale. We analyze this method using a dataset of micro-ultrasound acquired from 578 patients who underwent prostate biopsy, and compare our model to baseline models and other large-scale studies in the literature.

\emph{Results and Conclusions:} Our model shows consistent and substantial performance improvements compared to ROI-scale-only models.  It achieves $80.3\%$ AUROC, a statistically significant improvement over ROI-scale classification. We also compare our method to large studies on prostate cancer detection, using other imaging modalities. Our code is publicly available at \url{www.github.com/med-i-lab/TRUSFormer}.


}


\maketitle

\section{Introduction}\label{sec:intro}

Prostate cancer (PCa) is the second most common cancer in men worldwide. Early and accurate diagnosis and staging greatly increases the chances of successful treatment. The standard method for diagnosing and grading PCa is histopathological analysis of tissue samples of the prostate, typically obtained via trans-rectal ultrasound (TRUS) guided core biopsy. Because of the low sensitivity of conventional ultrasound in identifying lesions in the prostate~\cite{smeenge2012current}, TRUS-guided biopsy is generally \emph{systematic}, meaning that a large number of biopsy cores (typically 10-12) are uniformly sampled from the prostate rather than targeted to suspicious tissue locations. Systematic biopsy has a high false negative rate~\cite{ahmed2017diagnostic} and carries considerable risks of biopsy-related adverse effects due to the large number of samples that must be obtained~\cite{madej2012complication}. 

The emerging state-of-the-art in targeted prostate biopsy pairs ultrasound with multi-parametric MRI (mp-MRI). Compared to ultrasound, mp-MRI has significantly higher sensitivity in detecting cancerous lesions~\cite{ahmed2017diagnostic}, and, in combination with ultrasound, it can be used to identify and sample biopsy targets~\cite{siddiqui2013magnetic}. However, this requires pre-procedure MRI imaging, which is generally only available at large, well-funded urban centers, and registration of MRI to US images, which can be laborious and error-prone. 
 Therefore, the holy grail for cancer detection is to directly identify biopsy targets via ultrasound.

To improve ultrasound-based PCa detection, different ultrasound-based imaging modalities (e.g. shear wave elastography~\cite{salomon2008evaluation}, quantitative ultrasound~\cite{oelze2016review}, Doppler ultrasound~\cite{kelly1993prostate}, temporal-enhanced ultrasound~\cite{moradi2008augmenting}, micro-ultrasound~\cite{ghai2016assessing}, and contrast-enhanced ultrasound~\cite{mannaerts2020detection}) have been used. These methods, together with machine learning, have shown promise in interpreting complex ultrasound/tissue interactions. Classical machine learning models have been successfully used in conjunction with hand-selected features such as quantitative ultrasound~\cite{rohrbach2018high}, or shear-wave elastography~\cite{secasan2022artificial} coefficients. Despite promising results, manual feature selection is inherently limited to a small number and may miss complex properties that correlate with PCa. An alternative to classical machine learning methods are deep networks. Deep learning models have been proposed for PCa detection, for example from temporal-enhanced ultrasound ultrasound~\cite{fooladgar2022uncertainty}, micro-ultrasound~\cite{shao2020improving,gilany2022towards,wilson2022self}, and contrast-enhanced ultrasound~\cite{feng2018deep}. These studies generally formulate the problem of cancer detection as classification, where small regions of interest (ROIs) of an ultrasound image are classified into categories (benign or cancer). A notable exception is~\cite{javadi2020multiple}, which aims to classify \emph{biopsy cores} using temporal-enhanced ultrasound but only achieves $68\%$ AUROC. Classification of PCa at ROI-scale is limiting for two reasons: i) ground-truth histopathology annotations (i.e. Gleason Scores) describe tissue properties at the scale of the entire biopsy core.  Therefore, these only approximate the spread of cancer at the smaller ROI scale. Such weak labeling makes it challenging to train and evaluate robust ROI-scale models; 
 ii) ROI-scale models only take into account individual ROIs and cannot take advantage of contextual information, such as the properties of surrounding tissue or larger scale tissue patterns, when identifying cancer. 
Context is important in identifying PCa: for example, for histopathology labeling, PCa grade is determined from fine-scale features (e.g., degree of cell differentiation), medium-scale features (e.g., gland shape, size and uniformity, and infiltration patterns), and coarse-scale features (e.g., primary vs. secondary tumor types in the tissue sample)~\cite{gordetsky2016grading}.




To address the limitations of conventional ROI-scale models, we propose a multiscale approach (ROI-scale \emph{and} core-scale) to PCa detection. At the ROI scale, following the successful approach of~\cite{wilson2022self}, we use a self-supervised convolutional network to extract features from ROIs of raw backscattered ultrasound data. At the core-scale, inspired by successful methods in digital pathology (e.g~\cite{campanella2019clinical}), we treat a core as a bag of ROIs and use multiple instance learning (MIL) to aggregate ROI features and predict tissue type for the core. Specifically, we use an attention-based transformer~\cite{vaswani2017attention} model to aggregate ROI-level features, similar to attention-based MIL~\cite{ilse2018attention}.    
%
As a by-product, the attention maps produced by the transformer model allow us to identify the ROIs that contribute most to the prediction of tissue types in a biopsy core, hence localizing the tissue types at the ROI-scale. 

Using a large multi-center dataset of micro-ultrasound data, we demonstrate that our method significantly outperforms baseline approaches in distinguishing cancerous from benign cores. We demonstrate the additional explainability offered by the model via the attention maps. Furthermore, when visualizing the latent space of our multi-scale model,  the data is not only clustered as cancer and benign but also distributed according to the severity of cancer (Gleason score). 
To the best of our knowledge, this is the first method for micro-ultrasound based PCa detection that combines ROI-scale \emph{and} core-scale analyses.
\section{Materials}

\emph{Data Acquisition}: We use private data from 578 patients, from five centers, who underwent systematic TRUS-guided biopsy as part of a clinical trial (clinicaltrials.gov NCT02079025) and after agreements are provided by patients. Biopsies are performed with an ExactVu micro-ultrasound~\cite{rohrbach2018high} system (Exact Imaging, Markham, Canada). Prior to firing the biopsy gun, a single radio-frequency (RF) ultrasound image is recorded with 28~\textit{mm} depth and 46~\textit{mm} width in the sagittal plane. Images are paired with the histopathology findings for the corresponding biopsy core. For most patients, 10-12 biopsy cores are obtained. Among the 6607 total cores, 5727 are benign (defined as GS6 or lower), 675 are GS7, 134 are GS9, 60 are GS9, and 11 are GS10.\\
\emph{Data Selection:} First, the data of each center is divided to three patient-exclusive training, validation, and test sets with proportion 60\%-15\%-25\%, respectively. Then, the patients of all centers are grouped into an overall train, validation, and test cohort. For performing cross-validation, different validations are selected from training set. Additionally following previous works~~\cite{shao2020improving,rohrbach2018high}, cores with cancer involvement~$\leq 40\%$~of the core tissue are excluded in all datasets, and then benign cores are randomly selected to match the number of cancerous cores. Altogether, the training, cross-validation, and test sets include 864, 219, and 315 cores, respectively. 



\noindent\emph{Data Preprocessing}: For each biopsy core, the needle trace region is identified on the corresponding ultrasound image. From the needle trace region, several overlapping ROIs, corresponding to tissue areas of size 5~\textit{mm} by 5~\textit{mm} are automatically selected, moving in 1~\textit{mm} by 1~\textit{mm} ``strides". Following previous work \cite{wilson2022self}, an ROI is considered inside the needle trace region if it overlaps with the region by at least 66~\%. This selection process generates a bag of 55 ROIs per core. ROIs are resized from 1780 by 55 to 256 by 256 pixels by upsampling in the lateral and downsampling\footnote{The RF is bandlimited in the axial direction such that no loss of information occurs by downsampling.} in the axial directions. ROIs are instance-normalized by truncating outlier pixels ($>4$ standard deviations from the mean for that image) and rescaling intensity to the range of (0, 1).

\section{Methods}\label{sec:methods}

\begin{figure}
    \centering
    \includegraphics[scale=0.38]{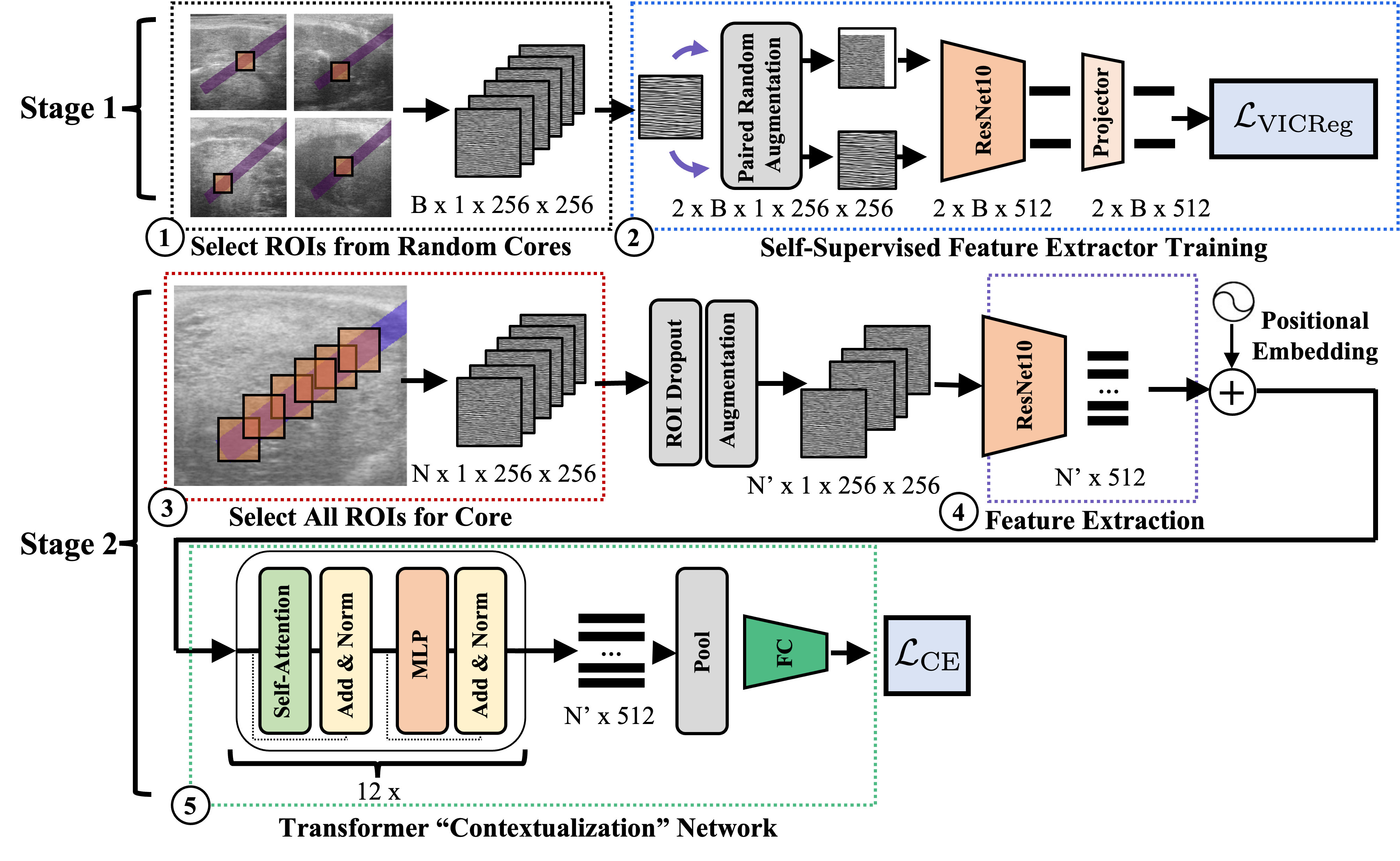}
    \caption{Summary of the proposed method. In Stage 1, batches of RF ultrasound ROIs are randomly selected from different cores (1) and used to train a feature extractor using the VICReg \cite{bardes2021vicreg} method (2). In Stage 2, ROIs for the cores are concatenated together (3) and features are extracted using the pretrained feature extractor (4). Features are passed to the transformer network that uses self-attention to create a contextualized global representation of the core, from which it predicts the class (cancer or benign) (5).}
    \label{fig:my_label}
\end{figure}

Our proposed method consists of two main parts (Fig. \ref{fig:my_label}). First, we use self-supervised learning (SSL) to train a convolutional neural network to extract informative feature representations for RF ultrasound ROIs. Second, we utilize MIL: we treat a biopsy core as a bag of ROIs, extract a bag of features from the ROIs using the pretrained network, and use an aggregation network to predict tissue type from the bag of features. The aggregator network uses attention to gather context from the various input ROIs, allowing it to identify cancer more robustly than looking at individual ROIs in isolation. 

\subsection{Self-supervised Feature Extraction}
Following \cite{wilson2022self}, we use the Variance-Invariance-Covariance Regularization (VICReg)~\cite{bardes2021vicreg} SSL method to train a modified ResNet architecture as the backbone network to extract 512 features to represent each ultrasound ROI. 
 
VICReg works as follows: Given an extracted RF ultrasound ROI $x_i^{\text{raw}}$, two augmentations $t$ and $t'$ are sampled from a distribution $\tau$, and applied to the ROI to produce two augmented ``views", $x_i=t(x_i^{\text{raw}})$ and $x'_i=t'(x_i^{\text{raw}})$. Augmentations for RF ultrasound are studied in \cite{wilson2022self}.  Different from that work, we use a strategy similar to \cite{bardes2021vicreg}, i.e., composition of random crops (scale 0.5-1.0), and horizontal and vertical flips, which we found to work well. 
The augmented data are then mapped to representation vectors $h_i$ and $h'_i$ using the backbone network $h_i=f_\theta(x_i)$. Then, the representations are projected to $z_i$ and $z'_i$ using the projection network $z_i=g_\phi(h_i)$. 

During self-supervised training, batches $Z$ and $Z'$ are collected from $z_i$ and $z_i'$, respectively (i.e. $Z = \verb|concat| (\{z_i : i \in \verb|batch_indices|  \})$). The weights $\theta$ and $\phi$ are adjusted to minimize the loss of VICReg network, $\ell(Z, Z')$. This loss is the weighted sum of three regularization terms, i.e. invariance $s(Z, Z')$, variance $v(Z)$, and covariance $c(Z)$ losses defined as follows:  
\begin{align}\label{eqn:vicreg_loss}
    \ell(Z, Z') & = \lambda s(Z,Z') + \mu [v(Z)+v(Z')] + \nu [c(Z)+c(Z')] ,
\end{align}
where the weights $\lambda$, $\mu$, and $\nu$ are hyperparameters (following \cite{bardes2021vicreg}, we used 25, 25, 1, which worked well). 
The invariance loss $s(Z,Z')$ is the mean square error loss, maximizing the agreement between projection vectors and ensuring that only related information is captured. The variance loss $\nu(Z)$ acts to maintain the variance of features across batches, avoiding the collapse of the representation. The covariance loss $c(Z)$ minimizes feature redundancy by minimizing cross-correlations projected features.

\subsection{Context-Aware Core Classification}

The second stage of our method is to train a biopsy-core classification network. The input to this network is the bag of ROIs from the needle region of a core, stacked along the first axis, denoted by $X$ ($X \in \mathbb{R}^{n \times 256 \times 256}$, where $n$ is the number of ROIs within the core). The output of the network is a two-dimensional class score $\hat{y} \in \mathbb{R}^2$. The network first applies the pre-trained feature extractor $f_\theta$ from the previous stage to each ROI, obtaining the bag of features $Z \in \mathbb{R}^{n \times 512}$. This is passed to the feature aggregator model $g_\phi$ to obtain class scores. During training, a dropout layer randomly drops some ROIs from the core to avoid seeing the same bag for the core each time.

The feature aggregator $g_\phi$ is a self-attention model based on a Transformer architecture~\cite{vaswani2017attention} with 12 attention blocks, 8 heads, inner dimension 256 and MLP dimension 512\footnote{We empirically found this to be the best configuration after a manual search through: layers 1, 8, 12, 16, inner dimensions 64, 128, 256, 768, and heads 4, 8.}. The model uses learnable positional embeddings, which are added to the input features to make the model aware of the relative positions of ROIs within the core. The model also includes a linear layer which projects the 512 input dimensions to a smaller dimension $d$ (we use $d=256$). The main body is a transformer encoder consisting of alternating multi-head self-attention layers and multi-layer perceptrons. A single self-attention head with input $Y \in \mathbb{R}^{n \times d}$ and output $Y'\in \mathbb{R}^{n\times d}$ uses the rule:
\begin{equation}
    Y' = A V; \quad A = \text{softmax}(\frac{QK^T}{\sqrt{d}}); 
\end{equation}
\begin{equation}
    Q = W^Q Y; K=W^K Y; V=W^V Y,
\end{equation}
for the learnable weights $W^Q$, $W^K$, and $W^V$ ($A$ is the so-called ``attention" matrix). Self-attention blocks combine information about each ROI with relevant information about other ROIs in bag at each layer, gathering context into the output state which is pooled and classified using a linear layer.

\subsubsection{Attention Mapping}\label{atten_map}
Our model takes as input a sequence of ultrasound image ROIs and outputs a class prediction. We use the technique of Chefer et al.~\cite{chefer2021generic} to determine the extent to which each ROI contributes to either the ``benign" or ``cancerous" class of a core. Given an input sequence of length $n$, the relevancy map $R$ is initialized as an identity matrix $I_n$, and  propagated through the attention layers via the update rule $R = R + \bar{A} R$.
Here, $\bar{A}$ can be thought of as a modified attention matrix that measures only contributions to a desired class. It is given by:
\begin{equation}
    \bar{A} = E_h[(A \odot \nabla A)^+], 
\end{equation}
where $\nabla$ denotes the gradient with respect to a class score (cancer or benign), $\odot$ denotes the Hadamard product and $E_h$ denotes the mean of different attention heads in a layer. At the last layer, the relevancy map is pooled to an $n \times 1$ vector whose $i$'th entry is the relative contribution of the $i$'th ROI for this class. 


\subsection{Experiments}

\emph{Stage 1}: We train the feature extractor for 200 epochs using a batch size of 64 and the Adam optimizer, warming up the learning rate for 10 epochs to $1\text{e}-4$ and cosine annealing (Schedule A). The model is assessed, as it trains, by ``online evaluation", periodically fitting a linear classifier to predict a class label (cancer or benign) for each ROI based on the features generated by the model. The best feature extractor is chosen based on online evaluation performance on the validation set. \emph{Stage 2}: the core classification and feature extractor models are trained end-to-end for 70 epochs using the Adam optimizer. The best model is also chosen based on validation performance tested after each epoch. The learning rate (LR) for the transformer network is warmed up over 5 epochs to 1e-4 and cosine annealed. The LR for the feature extractor is warmed up over 10 epochs to 3e-5 and cosine annealed.

We compared TRUSFormer to several other methods which we implemented, trained and tested on our dataset. First, ROI-scale approaches: (a) end-to-end supervised learning using the same ResNet architecture as TRUSformer's feature extractor, (b) the method of \cite{gilany2022towards} which enhanced supervised learning with methods to handle weak labels and uncertainty quantification, (c) the method of \cite{wilson2022self} which used SSL pretraining followed by supervised ROI-scale finetuning. As in these studies, ROI predictions were converted to biopsy core predictions by taking the average of ROI predictions across the set of ROIs inside the core. Second, we compared our method to attention-based MIL \cite{ilse2018attention} which is another example of a core-scale approach, and similar to our approach, aggregates ROI feature embeddings using an attention mechanism, but does not use the SSL pretraining step. Metrics are reported for the test data as average~$+/-$~standard deviation across 16 runs. 





\section{Results}\label{sec4}

Table \ref{tb:comp_w_bases} summarizes the core classification performance of TRUSFormer compared to other methods implemented on our dataset\footnote{Performance is reported across all data centers. No statistically significant differences in performance between different data centers were observed.}. 
It has the best performance in AUROC and Average Precision, with a considerable improvement over fully supervised ROI-scale methods ($+4\%$ AUROC, $+5\%$ Average Precision), large improvement over fully supervised core-scale methods ($+15\%$ AUROC, $+16\%$ Average Precision), and small but statistically significant\footnote{two-tailed p-value $<0.05$} ($+2\%$ AUROC, $+2.5\%$ Average Precision) improvement over SSL methods. We conclude that (1) improvements in performance are due both to the introduction of SSL feature extraction from ROIs \emph{and} the core-scale MIL, and (2) SSL seems necessary for core-scale methods to work, likely because good ROI features cannot be learned from scratch using core-scale MIL. Although TRUSformer has worse specificity than some methods, this could be improved by raising the detection threshold (at the cost of lowering sensitivity).
\begin{table}[th!]
  \caption{Comparison of the proposed method (TRUSformer) with other proposed models for PCa detection. Metrics are based on performance in classifying benign vs. cancerous (Gleason score 7 and above) biopsy cores. For ROI-scale classifiers, the core class prediction is the average of ROI for the core.
  }
  \label{tb:comp_w_bases}
  \centering
  \begin{tabular}{l c c c c}
    \toprule
    \textbf{Method} & AUROC & Avg-Prec & Sens & Spec\\
    \midrule
    
    \textbf{{ROI-scale classifier:}}&&&&\\
    
    \multirow{1}{*}{Supervised} & 76.3\scriptsize{$\pm$1.7} &  73.7\scriptsize{$\pm$2.2} & 63.6\scriptsize{$\pm$12.9} & \textbf{71.1}\scriptsize{$\pm$7.7}\\
    
    \multirow{1}{*}{{EDL + Coteaching \cite{gilany2022towards}}} & 75.9\scriptsize{$\pm$0.5} &  72.6\scriptsize{$\pm$0.1} & 63.3\scriptsize{$\pm$7.5} & 70.8\scriptsize{$\pm$5.9}\\

    \multirow{1}{*}{SSL + Linear Eval {\cite{wilson2022self}}} & 77.9\scriptsize{$\pm$1.1} &  74.8\scriptsize{$\pm$1.0} & 75.6\scriptsize{$\pm$4.3} & 66.8\scriptsize{$\pm$3.9}\\

    \multirow{1}{*}{SSL + Finetuning {\cite{wilson2022self}}} &  78.2\scriptsize{$\pm$3.8} &  75.2\scriptsize{$\pm$4.4} & \textbf{90.9}\scriptsize{$\pm$7.0} &41.3\scriptsize{$\pm$18.8}\\
    
    \midrule
    \textbf{{Core-scale classifier:}}&&&&\\
    
    \multirow{1}{*}{{Attention MIL \cite{ilse2018attention}}} & 63.0\scriptsize{$\pm$3.2} &  60.4\scriptsize{$\pm$2.5} & 60.0\scriptsize{$\pm$3.6} & 59.9\scriptsize{$\pm$3.4}\\
    
    \multirow{1}{*}{{Gated Attention MIL \cite{ilse2018attention}}} & 65.0\scriptsize{$\pm$2.5} &  62.2\scriptsize{$\pm$2.9} & 56.7\scriptsize{$\pm$1.9} & 67.9\scriptsize{$\pm$4.1}\\
    
    \multirow{1}{*}{TRUSformer (ours)} & \textbf{80.3}\scriptsize{$\pm$2.0} &  \textbf{78.7}\scriptsize{$\pm$1.9} & 88.0\scriptsize{$\pm$11.6} & 51.2\scriptsize{$\pm$14.6}\\

    \bottomrule

  \end{tabular}
\end{table}

    


    



\begin{figure}
    \centering
    \includegraphics[scale=.30]{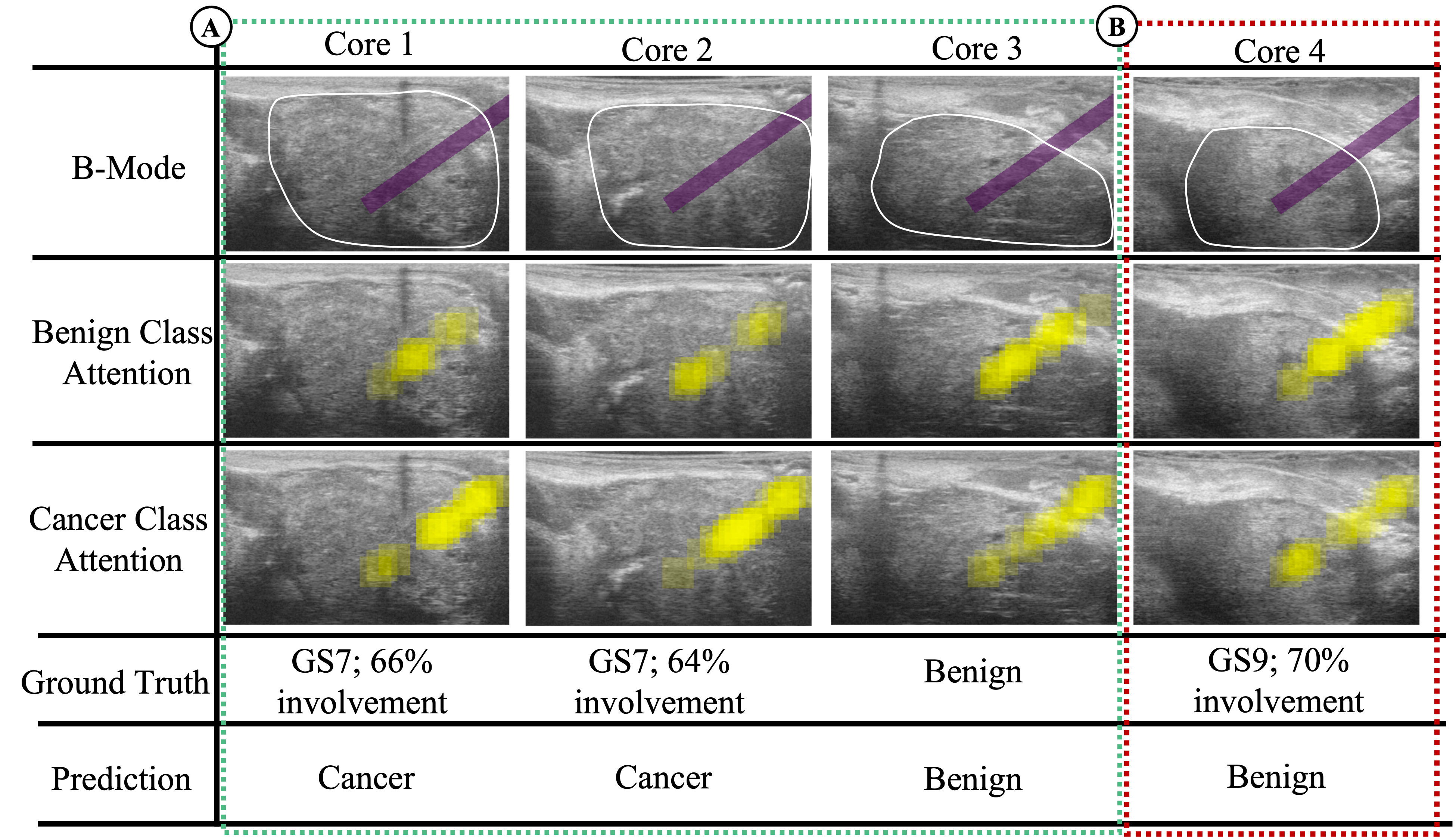}
    \caption{Qualitative analysis of model predictions and attention maps. The purple box on B-mode images in the first row show the biopsy core locations, and the white line shows the prostate outlines. For four cores, we compare TRUSformer's prediction (bottom row) for the core to ground truth (second row from bottom). We show attention maps produced by the model to see which ROIs in the core contribute to the ``benign" (second row from top) vs. ``cancer" (third row from top) class predictions. The left three cores (A) are from a patient for whom the model was consistently accurate, whereas the final core (B) is a failure case.}
    \label{fig:attn}
\end{figure}

Figure \ref{fig:attn} depicts visualizations of the model output. For 4 cores, we compare TRUSformer's prediction (bottom row) for the core to the ground truth (second row from bottom). We show attention maps produced by the model to see which ROIs in the core contribute to the ``benign" (second row from top) vs. ``cancer" (third row from top) class predictions. Cores 1-3 belong to one patient (A)  and core 4 belongs to another patient (B). The model was consistently accurate for cores of patient A. For cores 1 and 2, the attention values for cancer are dominant compared to the attention values for benign, and the ratio of benign to cancer attention areas approximately correspond to the involvement of cancer in the cores. For core 3, benign attention was higher in the prostate region, whereas cancer attention appeared higher outside the prostate. The final core from patient B is a failure case: this core was incorrectly predicted as benign, and the model appears primarily attentive to the top right corner (outside the prostate region). Since non-prostatic tissue can neither identify nor exclude cancer, excessive attention to this region may explain the model's failure here. 
Because the model sees fewer non-prostatic ROIs during training, features of these patches may appear as outliers and draw model attention. This could be improved with future work by adding more non-prostatic ROIs in training.

\begin{figure}
    \centering
    \includegraphics[scale=.07]{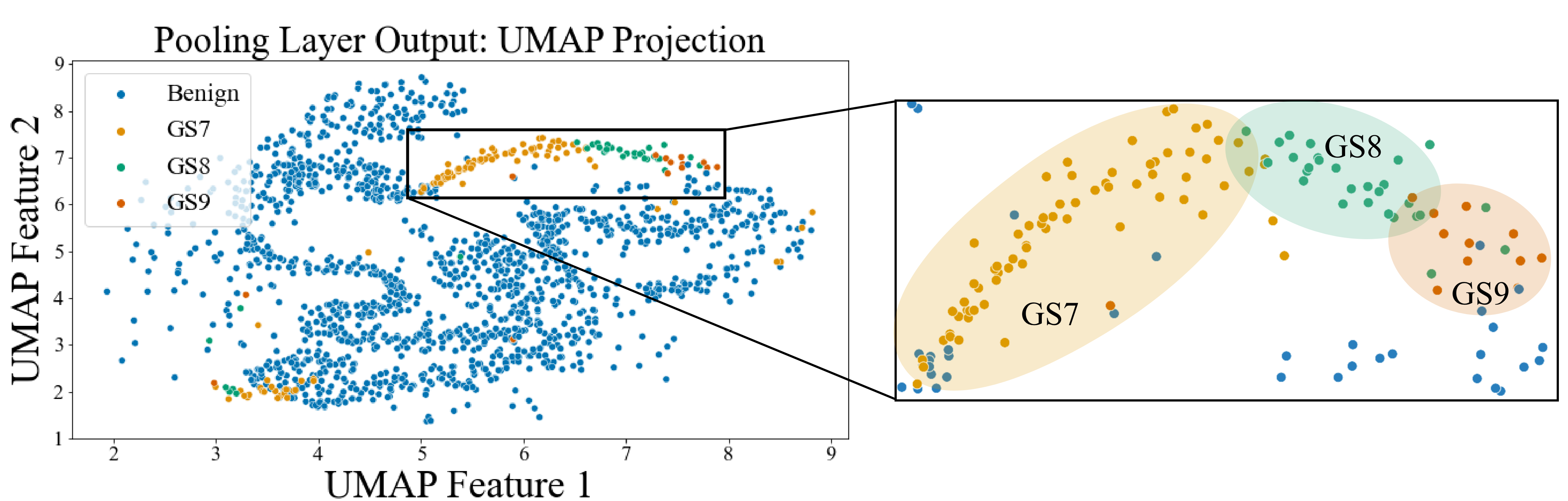}
    \caption{TRUSformer pooling layer features are visualized for each core in test set via UMAP.
    Despite not using the Gleason score labels in training, the UMAP representations show clusters of different Gleason scores.
    }
    \label{fig:umap}
\end{figure}

Figure \ref{fig:umap} visualizes the feature space of TRUSformer's second last layer (the pooling layer just before the classification layer) by projecting the features for all of the cores in the test set to two dimensions using UMAP~\cite{mcinnes2018umap}. Although the model was trained for binary classification (cancer vs. benign), we see apparent clustering of the cores according to their Gleason score. As Gleason scores partly depend on the distribution of tissue types throughout the core (not just on isolated ROIs), we hypothesize that the ability to see clusters with different Gleason scores is due to the core-scale nature of our model. 

Table \ref{tb:comp_w_literature} summarizes our method to other imaging methods (note the limitations of this comparison due to differences in data, cross-validation methods, and performance reporting). Our method compares favorably with other micro-ultrasound methods, achieving similar sensitivity but much better specificity than PRIMUS~\cite{ghai2016assessing} and comparable performance to QUS~\cite{rohrbach2018high}, despite the latter method requiring calibration to obtain the transducer point-spread-function using a phantom. Our model has the same sensitivity, but higher specificity than mp-MRI performance reported by the PROMIS clinical trial~\cite{ahmed2017diagnostic}. Since MRI fusion biopsy is the current state-of-the-art for PCa detection, matching the performance of mp-MRI means that this model has the potential for allowing targeted biopsy with ultrasound alone. However, the method still has limited specificity, and cannot confirm a cancer case without the need for pathology (from biopsy). 

\begin{table*}[th!]
  \caption{PCa detection performance reported in literature for various imaging modalities, compared to our approach. Performances are reported for cancer with Gleason score$\geq$7.}
  \label{tb:comp_w_literature}
  \centering
  \begin{tabular}{l c c c c c}
    \toprule
    \textbf{Method} & No. Patients & AUROC & Sens & Spec\\
    \midrule
    
    
  
    \multirow{1}{*}{PROMIS: TRUS biopsy \cite{ahmed2017diagnostic}} & 576 & - & 48 & 99 \\ 
    \multirow{1}{*}{PROMIS: PI-RADS + mp-MRI \cite{ahmed2017diagnostic}} & 576 &  - &  88 & 45\\
    


    \multirow{1}{*}{PRIMUS + micro-ultrasound \cite{ghai2016assessing}} & 93 & 60 & 80 & 37 \\ 
    \multirow{1}{*}{QUS + micro-ultrasound \cite{rohrbach2018high}} & 75 & 79 & 85 & 56 \\ 
    \multirow{1}{*}{TRUSformer + micro-US (ours)} & 125 (test set) & 80 & 88 & 51 &\\
    
    \bottomrule

  \end{tabular}
\end{table*}

\section{Conclusion}
We proposed TRUSformer, a multi-scale and context-aware method for PCa detection from trans-rectal ultrasound. TRUSformer improves PCa detection by eliminating the weak-labeling issue of ROI-scale models, and allowing the model to analyze multiple ROIs together in context to more robustly identify PCa. Although it was not trained explicitly to do so, TRUSformer appears to identify biopsy core-scale features that correlate well with distinct Gleason scores. Future work should further test TRUSFormer's generalization performance for cancer detection and grading via prospective studies.

\backmatter





\section*{Declarations}

We acknowledge Natural Sciences and Engineering Research Council of Canada,  and the Canadian Institute of Health Research. PM is supported by the CIFAR AI Chair and the Vector Institute.  
\bibliography{mybib_truncated}


\begin{thebibliography}{25}
\ifx \bisbn   \undefined \def \bisbn  #1{ISBN #1}\fi
\ifx \binits  \undefined \def \binits#1{#1}\fi
\ifx \bauthor  \undefined \def \bauthor#1{#1}\fi
\ifx \batitle  \undefined \def \batitle#1{#1}\fi
\ifx \bjtitle  \undefined \def \bjtitle#1{#1}\fi
\ifx \bvolume  \undefined \def \bvolume#1{\textbf{#1}}\fi
\ifx \byear  \undefined \def \byear#1{#1}\fi
\ifx \bissue  \undefined \def \bissue#1{#1}\fi
\ifx \bfpage  \undefined \def \bfpage#1{#1}\fi
\ifx \blpage  \undefined \def \blpage #1{#1}\fi
\ifx \burl  \undefined \def \burl#1{\textsf{#1}}\fi
\ifx \doiurl  \undefined \def \doiurl#1{\url{https://doi.org/#1}}\fi
\ifx \betal  \undefined \def \betal{\textit{et al.}}\fi
\ifx \binstitute  \undefined \def \binstitute#1{#1}\fi
\ifx \binstitutionaled  \undefined \def \binstitutionaled#1{#1}\fi
\ifx \bctitle  \undefined \def \bctitle#1{#1}\fi
\ifx \beditor  \undefined \def \beditor#1{#1}\fi
\ifx \bpublisher  \undefined \def \bpublisher#1{#1}\fi
\ifx \bbtitle  \undefined \def \bbtitle#1{#1}\fi
\ifx \bedition  \undefined \def \bedition#1{#1}\fi
\ifx \bseriesno  \undefined \def \bseriesno#1{#1}\fi
\ifx \blocation  \undefined \def \blocation#1{#1}\fi
\ifx \bsertitle  \undefined \def \bsertitle#1{#1}\fi
\ifx \bsnm \undefined \def \bsnm#1{#1}\fi
\ifx \bsuffix \undefined \def \bsuffix#1{#1}\fi
\ifx \bparticle \undefined \def \bparticle#1{#1}\fi
\ifx \barticle \undefined \def \barticle#1{#1}\fi
\bibcommenthead
\ifx \bconfdate \undefined \def \bconfdate #1{#1}\fi
\ifx \botherref \undefined \def \botherref #1{#1}\fi
\ifx \url \undefined \def \url#1{\textsf{#1}}\fi
\ifx \bchapter \undefined \def \bchapter#1{#1}\fi
\ifx \bbook \undefined \def \bbook#1{#1}\fi
\ifx \bcomment \undefined \def \bcomment#1{#1}\fi
\ifx \oauthor \undefined \def \oauthor#1{#1}\fi
\ifx \citeauthoryear \undefined \def \citeauthoryear#1{#1}\fi
\ifx \endbibitem  \undefined \def \endbibitem {}\fi
\ifx \bconflocation  \undefined \def \bconflocation#1{#1}\fi
\ifx \arxivurl  \undefined \def \arxivurl#1{\textsf{#1}}\fi
\csname PreBibitemsHook\endcsname

\bibitem{smeenge2012current}
\begin{barticle}
\bauthor{\bsnm{Smeenge}, \binits{M.}},
\bauthor{\bparticle{de~la} \bsnm{Rosette}, \binits{J.J.}},
\bauthor{\bsnm{Wijkstra}, \binits{H.}}:
\batitle{Current status of transrectal ultrasound techniques in prostate
  cancer}.
\bjtitle{Current opinion in urology}
\bvolume{22}(\bissue{4}),
\bfpage{297}--\blpage{302}
(\byear{2012})
\end{barticle}
\endbibitem

\bibitem{ahmed2017diagnostic}
\begin{barticle}
\bauthor{\bsnm{Ahmed}, \binits{H.U.}}, \betal:
\batitle{Diagnostic accuracy of multi-parametric mri and trus biopsy in
  prostate cancer (promis): a paired validating confirmatory study}.
\bjtitle{The Lancet}
\bvolume{389}(\bissue{10071}),
\bfpage{815}--\blpage{822}
(\byear{2017})
\end{barticle}
\endbibitem

\bibitem{madej2012complication}
\begin{barticle}
\bauthor{\bsnm{Madej}, \binits{A.}},
\bauthor{\bsnm{Wilkosz}, \binits{J.}},
\bauthor{\bsnm{R{\'o}{\.z}a{\'n}ski}, \binits{W.}},
\bauthor{\bsnm{Lipi{\'n}ski}, \binits{M.}}:
\batitle{Complication rates after prostate biopsy according to the number of
  sampled cores}.
\bjtitle{Central European journal of urology}
\bvolume{65}(\bissue{3}),
\bfpage{116}
(\byear{2012})
\end{barticle}
\endbibitem

\bibitem{siddiqui2013magnetic}
\begin{barticle}
\bauthor{\bsnm{Siddiqui}, \binits{M.M.}}, \betal:
\batitle{Magnetic resonance imaging/ultrasound--fusion biopsy significantly
  upgrades prostate cancer versus systematic 12-core transrectal ultrasound
  biopsy}.
\bjtitle{European urology}
\bvolume{64}(\bissue{5}),
\bfpage{713}--\blpage{719}
(\byear{2013})
\end{barticle}
\endbibitem

\bibitem{salomon2008evaluation}
\begin{barticle}
\bauthor{\bsnm{Salomon}, \binits{G.}}, \betal:
\batitle{Evaluation of prostate cancer detection with ultrasound real-time
  elastography: a comparison with step section pathological analysis after
  radical prostatectomy}.
\bjtitle{European urology}
\bvolume{54}(\bissue{6}),
\bfpage{1354}--\blpage{1362}
(\byear{2008})
\end{barticle}
\endbibitem

\bibitem{oelze2016review}
\begin{barticle}
\bauthor{\bsnm{Oelze}, \binits{M.L.}},
\bauthor{\bsnm{Mamou}, \binits{J.}}:
\batitle{Review of quantitative ultrasound: Envelope statistics and backscatter
  coefficient imaging and contributions to diagnostic ultrasound}.
\bjtitle{IEEE TUFFC}
\bvolume{63}(\bissue{2}),
\bfpage{336}--\blpage{351}
(\byear{2016})
\end{barticle}
\endbibitem

\bibitem{kelly1993prostate}
\begin{barticle}
\bauthor{\bsnm{Kelly}, \binits{I.}},
\bauthor{\bsnm{Lees}, \binits{W.}},
\bauthor{\bsnm{Rickards}, \binits{D.}}:
\batitle{Prostate cancer and the role of color doppler us.}
\bjtitle{Radiology}
\bvolume{189}(\bissue{1}),
\bfpage{153}--\blpage{156}
(\byear{1993})
\end{barticle}
\endbibitem

\bibitem{moradi2008augmenting}
\begin{barticle}
\bauthor{\bsnm{Moradi}, \binits{M.}}, \betal:
\batitle{Augmenting detection of prostate cancer in transrectal ultrasound
  images using svm and rf time series}.
\bjtitle{IEEE Transactions on Biomedical Engineering}
\bvolume{56}(\bissue{9}),
\bfpage{2214}--\blpage{2224}
(\byear{2008})
\end{barticle}
\endbibitem

\bibitem{ghai2016assessing}
\begin{barticle}
\bauthor{\bsnm{Ghai}, \binits{S.}}, \betal:
\batitle{Assessing cancer risk on novel 29 mhz micro-us images of the prostate:
  creation of the micro-us protocol for prostate risk identification}.
\bjtitle{The Journal of urology}
\bvolume{196}(\bissue{2}),
\bfpage{562}--\blpage{569}
(\byear{2016})
\end{barticle}
\endbibitem

\bibitem{mannaerts2020detection}
\begin{barticle}
\bauthor{\bsnm{Mannaerts}, \binits{C.K.}}, \betal:
\batitle{Detection of clinically significant prostate cancer in
  biopsy-na{\"\i}ve men: direct comparison of systematic biopsy,
  multiparametric mri-and contrast-ultrasound-dispersion imaging-targeted
  biopsy}.
\bjtitle{BJU international}
\bvolume{126}(\bissue{4}),
\bfpage{481}--\blpage{493}
(\byear{2020})
\end{barticle}
\endbibitem

\bibitem{rohrbach2018high}
\begin{barticle}
\bauthor{\bsnm{Rohrbach}, \binits{D.}}, \betal:
\batitle{High-frequency quantitative ultrasound for imaging pca using a novel
  micro-us scanner}.
\bjtitle{Ultrasound in medicine \& biology}
\bvolume{44}(\bissue{7}),
\bfpage{1341}--\blpage{1354}
(\byear{2018})
\end{barticle}
\endbibitem

\bibitem{secasan2022artificial}
\begin{barticle}
\bauthor{\bsnm{Secasan}, \binits{C.C.}}, \betal:
\batitle{Artificial intelligence system for predicting prostate cancer lesions
  from shear wave elastography measurements}.
\bjtitle{Current Oncology}
\bvolume{29}(\bissue{6}),
\bfpage{4212}--\blpage{4223}
(\byear{2022})
\end{barticle}
\endbibitem

\bibitem{fooladgar2022uncertainty}
\begin{bchapter}
\bauthor{\bsnm{Fooladgar}, \binits{F.}}, \betal:
\bctitle{Uncertainty-aware deep ensemble model for targeted ultrasound-guided
  prostate biopsy}.
In: \bbtitle{2022 IEEE 19th International Symposium on Biomedical Imaging
  (ISBI)},
pp. \bfpage{1}--\blpage{5}
(\byear{2022})
\end{bchapter}
\endbibitem

\bibitem{shao2020improving}
\begin{barticle}
\bauthor{\bsnm{Shao}, \binits{Y.}},
\bauthor{\bsnm{Wang}, \binits{J.}},
\bauthor{\bsnm{Wodlinger}, \binits{B.}},
\bauthor{\bsnm{Salcudean}, \binits{S.E.}}:
\batitle{Improving prostate cancer (pca) classification performance by using
  three-player minimax game to reduce data source heterogeneity}.
\bjtitle{IEEE Transactions on Medical Imaging}
\bvolume{39}(\bissue{10}),
\bfpage{3148}--\blpage{3158}
(\byear{2020})
\end{barticle}
\endbibitem

\bibitem{gilany2022towards}
\begin{bchapter}
\bauthor{\bsnm{Gilany}, \binits{M.}}, \betal:
\bctitle{Towards confident detection of prostate cancer using high resolution
  micro-ultrasound}.
In: \bbtitle{MICCAI 2022, Proceedings, Part IV},
pp. \bfpage{411}--\blpage{420}
(\byear{2022}).
\bcomment{Springer}
\end{bchapter}
\endbibitem

\bibitem{wilson2022self}
\begin{botherref}
\oauthor{\bsnm{Wilson}, \binits{P.F.}}, et al.:
Self-supervised learning with limited labeled data for prostate cancer
  detection in high frequency ultrasound.
arXiv preprint arXiv:2211.00527
(2022)
\end{botherref}
\endbibitem

\bibitem{feng2018deep}
\begin{barticle}
\bauthor{\bsnm{Feng}, \binits{Y.}}, \betal:
\batitle{A deep learning approach for targeted contrast-enhanced ultrasound
  based prostate cancer detection}.
\bjtitle{IEEE/ACM transactions on computational biology and bioinformatics}
\bvolume{16}(\bissue{6}),
\bfpage{1794}--\blpage{1801}
(\byear{2018})
\end{barticle}
\endbibitem

\bibitem{javadi2020multiple}
\begin{barticle}
\bauthor{\bsnm{Javadi}, \binits{G.}}, \betal:
\batitle{Multiple instance learning combined with label invariant synthetic
  data for guiding systematic prostate biopsy: a feasibility study}.
\bjtitle{IJCARS 2020}
\bvolume{15}(\bissue{6}),
\bfpage{1023}--\blpage{1031}
(\byear{2020})
\end{barticle}
\endbibitem

\bibitem{gordetsky2016grading}
\begin{barticle}
\bauthor{\bsnm{Gordetsky}, \binits{J.}},
\bauthor{\bsnm{Epstein}, \binits{J.}}:
\batitle{Grading of prostatic adenocarcinoma: current state and prognostic
  implications}.
\bjtitle{Diagnostic pathology}
\bvolume{11}(\bissue{1}),
\bfpage{1}--\blpage{8}
(\byear{2016})
\end{barticle}
\endbibitem

\bibitem{campanella2019clinical}
\begin{barticle}
\bauthor{\bsnm{Campanella}, \binits{G.}}, \betal:
\batitle{Clinical-grade computational pathology using weakly supervised deep
  learning on whole slide images}.
\bjtitle{Nature medicine}
\bvolume{25}(\bissue{8}),
\bfpage{1301}--\blpage{1309}
(\byear{2019})
\end{barticle}
\endbibitem

\bibitem{vaswani2017attention}
\begin{botherref}
\oauthor{\bsnm{Vaswani}, \binits{A.}}, et al.:
Attention is all you need.
Advances in neural information processing systems
\textbf{30}
(2017)
\end{botherref}
\endbibitem

\bibitem{ilse2018attention}
\begin{bchapter}
\bauthor{\bsnm{Ilse}, \binits{M.}},
\bauthor{\bsnm{Tomczak}, \binits{J.}},
\bauthor{\bsnm{Welling}, \binits{M.}}:
\bctitle{Attention-based deep multiple instance learning}.
In: \bbtitle{International Conference on Machine Learning},
pp. \bfpage{2127}--\blpage{2136}
(\byear{2018}).
\bcomment{PMLR}
\end{bchapter}
\endbibitem

\bibitem{bardes2021vicreg}
\begin{botherref}
\oauthor{\bsnm{Bardes}, \binits{A.}},
\oauthor{\bsnm{Ponce}, \binits{J.}},
\oauthor{\bsnm{LeCun}, \binits{Y.}}:
Vicreg: Variance-invariance-covariance regularization for self-supervised
  learning.
arXiv preprint arXiv:2105.04906
(2021)
\end{botherref}
\endbibitem

\bibitem{chefer2021generic}
\begin{bchapter}
\bauthor{\bsnm{Chefer}, \binits{H.}},
\bauthor{\bsnm{Gur}, \binits{S.}},
\bauthor{\bsnm{Wolf}, \binits{L.}}:
\bctitle{Generic attention-model explainability for interpreting bi-modal and
  encoder-decoder transformers}.
In: \bbtitle{ICCV},
pp. \bfpage{397}--\blpage{406}
(\byear{2021})
\end{bchapter}
\endbibitem

\bibitem{mcinnes2018umap}
\begin{botherref}
\oauthor{\bsnm{McInnes}, \binits{L.}},
\oauthor{\bsnm{Healy}, \binits{J.}},
\oauthor{\bsnm{Melville}, \binits{J.}}:
Umap: Uniform manifold approximation and projection for dimension reduction.
arXiv preprint arXiv:1802.03426
(2018)
\end{botherref}
\endbibitem

\end{thebibliography}


\end{document}